\documentstyle[12pt]{article}
\begin{document}
\baselineskip=24pt
\newcommand{\Lam}{\Lambda_{\scriptscriptstyle {\rm \overline{MS}} }}
\newcommand{\Lamm}[1]{\Lambda {\vspace{-2pt} \scriptscriptstyle ^{^{\rm #1}}
    _{\rm \overline{MS}} }}
\newcommand{\dsp}{\displaystyle}
\newcommand{\dfr}[2]{ \displaystyle\frac{#1}{#2} }
\newcommand{\Lag}{\Lambda \scriptscriptstyle _{ \rm GR} }
\newcommand{\pa}{p\parallel}
\newcommand{\pe}{p\perp}
\newcommand{\pet}{p\top}
\newcommand{\paa}{p'\parallel}
\newcommand{\pee}{p'\perp}
\newcommand{\pete}{p'\top}
\renewcommand{\baselinestretch}{1.5}
\begin{titlepage}
\vspace{-20ex}
\begin{flushright}
\vspace{-3.0ex}
    \it{AS-ITP-93-80} \\
\vspace{-2.0mm}
       \it{Dec., 1993}\\
\vspace{5.0ex}
\end{flushright}

\centerline{\Large 
The Instantaneous Approximation to the Transition Matrix Elements}
\centerline{\Large 
between Two Bound States \footnote{The Work was
supported in part by the National Natural Science Foundation of China
and the Grant LWTZ-1298 of Chinese Academy of Science.}}
\vspace{6.4ex}
\centerline{\large 
Chao-Hsi Chang$^{*,\dagger}$
and Yu-Qi Chen$^{\dagger}$}
\vspace{3.5ex}
\centerline{\sf $^*$CCAST (World Laboratory), P.O.Box 8730,
Beijing 100080, China}
\centerline{\sf $^\dagger$Institute of Theoretical Physics, Academia Sinica,
P.O.Box 2735, Beijing 100080, China\footnote{Mailing address.}}

\vspace{3ex}
\begin{center}
\begin{minipage}{5in}
\centerline{\large 
Abstract}
\vspace{1.5ex}
\small
{Under the framework of
the Bethe-Salpeter (B.S.) wave functions
and the Mandelstam formalism as well,
to make ``instantaneous approximation'' to a
transition matrix element (a current operator
sandwiched between two
bound-states of double heavy quarks)
is described. By taking the typical concerned decays as
examples, such as $B_c$ meson decaying to $J/\psi + (\bar l \nu)$,
the advantages of the approach and its limitations are illustrated.
Finally, potential applications to various processes for
possible double heavy flavoured systems, such as those of
$(Q' \bar Q)$ and $(Q' Q)$ ($m_Q, m_{Q'}\gg \Lambda_{QCD}$), are discussed.}
\end{minipage}
\end{center}

\end{titlepage}
\newpage

The Standard Model (S.M.), as well-known, has made great successful. The
flavour decays i.e. the weak decays, of the
fundamental fermions (quarks and leptons) are supposed to be
understood very well in the framework of the S.M. i.e. the vector
boson exchange and the CKM (Cabibbo-Kobayashi-Maskawa) mixing matrix
structure. However, except the leptons and the undiscovered top quark,
all the known quarks always decay as constituenets of hadrons, once
they are produced. This is due to the fact that colour
has confinement nature and the quarks carry
colour explicitly, thus a quark
always `fragment' into hadrons before decaying
once it is produced. (Only the undiscovered top quark may be an except,
as its mass is very great hence its lifetime is so
short that there is no time to fragment into hadrons before
decaying.) Thus we always see flavour decays through hadrons
in experiments and we have to deal with
the effects of the confinement when studying the flavour decays of quarks.
However, the confinement effects are of a nonperturbative nature
so very difficult to compute them reliablely.
Hence the weak decay problem has a very long story, which had started before
the S.M. was established. Nevertheless the progresses on the problem
have always being achieved steadly, especially
the recent achievements of the Heavy Effective Theory (HET) for one (or
two) light and one heavy flavour hadrons: $Q\bar q$ or
$Qqq$ ($m_Q>>m_q$)
and their antiparticles.
Due to the authors of HET$^{[1-6]}$, a great progress has
been achieved in understanding the weak decay
of heavy flavours under the approximation $m_Q\rightarrow \infty$
and with the corrections of the order $1/m_Q$ and higher. The authors note
at very beginning when they proposed the approach that
one of the applicable conditions of their formalism is that the mass of
the heavy quark in the concerned hadrons should be
greater than $\Lambda_{QCD}$
and much greater than that of the light one.

In the meantime as emphasized
by many authors[7-11], the $B_c$ meson will play a special role
for understanding the heavy flavour decays and some progresses
have achieved too.

The $B_c$ meson is the
ground state of the bound states for the system of the
heavy quark-antiquark pair $c$ and $\bar{b}$.
Both of the quarks $\bar{b}$ and $c$
are heavy quarks i.e. $m_Q\ll \Lambda_{QCD}, Q=b, c$.
Though both may be considered
being nonrelativistic,
their masses are compatible, thus the formalism$^{[1-6]}$ may
not be very appropriate, at least
careful examination are needed and considerable modifications are expected.

Of the three generations in the S.M.,
besides the well-studied systems of the $b \bar{b}$ and the $c \bar{c}$,
the quark-antiquark $c$ and $\bar{b}$
and their antiparticles are the only ones which still have
a long enough lifetime to form a bound state before
one of them decaying.
In contrary, the rest possible heavy quark pair systems
must contain a top or anti-top quark at least.
They will not have chance to form a bound state. It is
due to a great mass of the top quark ( the top mass $m_t$
greater than $140GeV^{[12]}$, though the top has
not discovered yet ) that causes a great phase space
so a very short lifetime even in the case only
weak decays are allowed, hence there is no time to
form a bound state before the top decaying.
The $c$ and $\bar{b}$ system so interesting
is attributed to the fact that
it is the only possible one which may make
mesons and the mesons have not been discovered yet.
Besides it plays
a special role in understanding the heavy flavour decays.
The $B_c$ meson, the ground state of the $c$ and $\bar{b}$ system,
not as that of the $c$(or $b$) and $\bar c$($\bar b$) systems,
carries flavors explicitly so no strong decay at all i.e.
the ground state is stable to the strong interaction.
It has a great advantage for studying the weak decay and
the bound state effects of a heavy quark-antiquark
system, even for the later only,
still better than those of the $J/\psi$ and $\Upsilon$ families.
Indeed, the ground state, the $B_c$ meson, has
many sizable and various weak decay channels, so
that it becomes possible to compare the results of
experiments and theories on the weak interaction effects
and the binding effects
of a heavy quark pair system widely.

Usually the above two kinds of effects in the meson decays
are divided into two categories from another aspect for
convenience: the short distance effects i.e.
the weak interaction responding for the flavour change
and its perturbative QCD corrections;
and the long distance ones i.e. the binding effects for the components.
In general speaking, although they are entangled,
the second ones, being of non-perturbative nature, are very difficult
to deal with aliably,
while the former ones, being perturbative
nature, are relatively easy to
treat by means of the Standard Model of electro-weak interaction and
perturbative QCD up to a desired accuracy. However, for the
non-perturbative binding effects of a heavy quark
system, great progresses
have been achieved under the framework of the QCD inspired potential model,
but in some phenomenological sense,
especially for the $J/\psi$ and $\Upsilon$ families.
Thus we may consult the potential model to treat the binding effects
for the $c$ and $\bar b$ system too.
According to the QCD inspired potential
model, the $B_c$ meson is very similar to the two families,
of $J/\psi$ and
$\Upsilon$, but has a different reduce mass only.
Thus the Shr\"odinger equation with the same QCD
inspired potential and a corresponding different reduce mass,
should be worked very well to depict the bending
effects of the $c$ and $\bar b$ system.
The Shr\"odinger equation, as known being in a sense of
nonrelativistics, hence to the static properties of the system, is
good enough, while to the decays, especially those with a great recoil
momentum may be problematic when using the solution(s) as the wave function(s)
directly. To find a way to use the Sch\"odinger wave function(s) properly
into the decays of the double heavy quark systems is the aim of the paper.

To our knowledge, the Mandelstam formalism[13] with the Bethe-Salpeter (B.S.)
wave functions is one of the best approaches to compute a matrix element
sandwiched between two single bound states, however
which is very often to be attributed at last when calculating
a whole decay matrix element.
Due to Salpeter[14], the
relation of a B.S. wave function to the Shr\"odinger one was established
quite long time ago by the so-called instantaneous approximation.
Therefore it seems no problem for the present purpose that
the Mandelstam formalism might be used with the help of the
Shr\"odinger wave functions by solving the reduced and corresponding
nonrelativistic equation. Nevertheless,
in the Salpeter's approach, the approximation taken at the centre mass system
of the bound state is implicated. If only
to apply the Shr\"odinger wave functions
to calculating those decay matrix elements sandwiched
between two states, but of which only one bound state is involved,
it is straightforward
relatively, provided one calculates such a matrix element at the centre
mass system of the bound state. However now for the most
important decays of the $B_c$ meson,
the problem is not so straightforward because
the decay matrix element is attributed to calculate a
weak current matrix element with quite a great momentum recoil,
the current operator sandwiched between two bound state.
In this paper we are generalizing the instantaneous approximation
to suit the interesting cases.

To be the start point, we suppose that
the transition matrix elements of the $B_c$ meson,
sandwiched between two single bound states,
( the form factors or the decay current matrix elements )
may be computed in a desired accuracy with the help of the Mandelstam
formalism no matter how great the momentum recoil, provided that
the one-particle irreducible green function, responsible
for the matrix element in the formalism,
matches to the trancated kernel of the B.S. equation well, as required by
the formalism$^{[15]}$. It is because the formalism is fully relativistic.
Now the problem is attributed to having reliable B.S. wave
functions for the quark bound states, appearing
in the formalism. In general as for hadrons the key problem to
obtain reliable B.S. wave functions is to establish the B.S.
equation with a reliable kernel and to find a suitable method
to solve the equation. However the nonperturbative nature
of the binding effects between the quarks from the QCD confinement
makes the problem very difficult.
For those matrix elements sandwiched between two
double heavy quark bound state(s) such as the $B_c$
meson and $J/\psi$ ( or $\Upsilon$ ) etc. the B.S.
wave functions occurring in the formalism
can be acquired reliablly by means of the
instantaneous approximation, i.e. the B.S. kernel for the double heavy
quark systems may be related to the QCD inspired potential of the
Sch\"odinger equation esteblished in a potential model framework, so the
B.S.'s and Schr\"odinger's wave functions are related too through the
approximation. Whereas one could not set the initial and final
states in a reference frame both at
rest for a real decay process, especially
when a matrix element with quite a great momentum recoil
is considered, thus the relation between the wave functions of the potential
model and the B.S. ones, established only at a rest frame (C.M.S.)
as indicated by the original instantaneous approximation, still cannot be
applied straightforward.
Here we constrain ourselves to consider the transitions between the
states of double heavy quark systems. The strategy, we adopt
is to generalize the instantaneous
approximation to apply to the whole matrix elements no matter how great
the momentum recoil. We note here that as the wave functions
are computed out only numerically in the potential
model, one cannot use the method by Lorentz boosting it directly
to pursue the present purpose reliably.
Our approach here, essentially to say, is
to calculate the matrix elements under the Mandelstam formalism
and to apply the `instantaneous approximation' to the matrix elements
in a whole, instead of that to the B.S.
equation only,
as done by Salpeter$^{[14]}$. One will see that with
our approach the calculations
on the matrix elements under the Mandelstam formalism and the B.S.
wave functions extracted from potential models as well,
are reliable even in the cases of the transitions with quite a great
momentum recoil. As previously discussed,
we develop the approach with the motivation from
the $B_c$ meson physics, thus we
will take the decays such as the $B_c$ meson to $J/\psi + (\bar l \nu)$
as examples to illustrate it in this paper.

To calculate the exclusive weak decays
of the $B_c$ meson, one needs to evaluate
the hadronic matrix elements, i.e. the weak current operator
sandwiched between the
initial state of the $B_c$ meson and the
concerned hadronic finial state.
We restrain ourselves to evaluate them in
the simplest cases, i.e. only those decays in their final state
having only one hadron
for semileptonic ones,
but two hadrons for nonleptonic ones.
In these cases, one may attribute the problem to evaluating
a matrix element of the weak current operator
sandwiched by two of single-hadron states
( for nonleptonic decays it is
due to the factorization assumption of calculating the decay amplitude ).

With the notation of a
weak charged current $J_\mu =V_\mu -A_\mu$, where $V_\mu$,
$A_\mu$ are the vector and the axial vector current respectively,
the matrix elements are related to the form
factors$^{[21,23]}$ as
\begin{equation}
\begin{array}{l}
< P(p') |V_\mu|B_c(p) > = f_+(p+p')_\mu + f_-(p-p')_\mu, \\[0.2in]
< V(p',\epsilon^\ast ) |V_\mu|B_c(p) >= i g\epsilon_{\mu\nu\rho\sigma}
           \epsilon^{\ast\nu}(p+p')^\rho(p-p')^\sigma,\\[0.2in]
< V(p',\epsilon^\ast ) |A_\mu|B_c(p) >=f \epsilon^{\ast}_{\mu}
       +a_+ (\epsilon^{\ast}\cdot p)(p+p')_\mu
       +a_- (\epsilon^{\ast}\cdot p)(p-p')_\mu.
\end{array}
\vspace{2mm}
\end{equation}
where $p$, $p'$ are the momenta of the $B_c$ and the outgoing hadron
respectively,
$P$ and $V$ denote the pseudoscalar and the vector
mesons respectively, $\epsilon$ is the polarization vector of the
vector meson. The form
factors are functions of Lorentz invariant variable $r^2\equiv (p-p')^2$.

In the literature one may find
two kinds of approaches to calculate these form factors.
One of them is BSW model$^{[16]}$,
in which the authors calculated the form factors at the maximum recoil
$r^2 = 0$ by means of the wave functions defined at the light cone system
under the quark model framework,
and then to extrapolate the result to all values
of $r^2$ by assuming the form factors dominated by a proper pole
of the nearest ones.
The other is IGSW model$^{[17]}$. The authors of Ref.[17]
calculated the form factors by using the wave functions
of the quark-model  (``mock meson'')
which treats the hadrons as a nonrelativistic object.
As argued by the authors, the approach is exactly valid in the limit of
weak binding and at the point of zero recoil.
However, in the cases with a large recoil, it is problematic.
For instance, for the decay
$B_c\rightarrow J/\psi +\rho$, i.e. the example we are considering,
although the initial state $B_c$
and the final
state $J/\psi$ both are of weak binding, the recoil of the
decay is not small so the IGSW model may not work very well.

However, inside the $B_c$ meson, both the $\bar{b}$ and $c$
are heavy so may be considered being nonrelativistic objects,
but their masses are comparable, thus the formalism$^{[1-6]}$ may
not be very appropriate, at least
careful examination and considerable modification are expected.
Furthermore, even though their approach is not suitable for
the $B_c$ meson, to establish a link
between the universal
Isgur-Wise function $\xi(v \cdot v')$ and the nonrelativistic
wave function of the heavy meson obtained
by the potential model would be still very interesting, where
$v$ and $v'$ are the four-velocity vectors of the meson
in the initial and final states respectively.

Now let us proceed to write down the matrix element with the help
of the Mandelstam formalism,
so to describe the approach explicitly.
It is known that the B-S equation of a fermion-antifermion bound
state takes the following form:
\begin{equation}
(\rlap/{p_1}-m_1)\;\chi_p(q)\;(\rlap/{p_2}+m_2)=i\int \dfr{d^4k}{(2\pi)^4}
\;V(p,k,q)\;\chi_p(k)
\vspace{2mm}\end{equation}
where $p_1$, and $p_2$ are the momenta of the constituent
particles 1 and 2 respectively.
They can be expressed in terms of the total and the relative momenta
$p$ and $q$ as
\begin{equation}
\begin{array}{cc}
p_1=\alpha_1p+q, &~~~~\alpha_1=\dfr{m_1}{m_1+m_2};\\[0.2in]
p_2=\alpha_2p-q, &~~~~\alpha_2=\dfr{m_2}{m_1+m_2},
\end{array}
\vspace{2mm}\end{equation}
$V(p,k,q)$ is the interaction kernel. It is well known that the B-S
wave function $\chi_p(q)$ satisfies the normalization condition:
\begin{equation}
\int\dfr{d^4q}{(2\pi)^4} \frac{d^4q'}{(2\pi)^4} {\rm tr}\{ \bar\chi_p(q)
\frac {\partial} {\partial p_0} [ S_1^{-1}(p_1) S_2^{-1}(p_2) \delta^4(q-q')
+V(p,q,q')]\chi_p(q')\}=2ip_0.
\vspace{2mm}\end{equation}

As for the decays caused by the quark $q_1$ of the meson,
according to the Mandelstam formalism[13] and the spectator
mechanism shown in Fig.1, the weak current matrix element
involving one hadron in the initial state and one
in the final state respectively, may be expressed
in terms of the B-S wave functions:
\begin{equation}
l^{\mu}(r)=i\int \dfr{d^4q}{(2\pi)^4}
{\rm tr}[ \bar{\chi}_p'(q')\Gamma^{\mu}_1\chi_{p}(q)
(\rlap/p_2+m_2)],
\vspace{2mm}\end{equation}
where $\chi_p(q)$, $\bar{\chi}_{p'}(q')$
are the B-S wave functions of the initial
state and the final state with the total momenta $p$, $p'$ and the relative
momenta $q$, $q'$ respectively;  $p_1$, $m_1$, $p'_1$, $m'_1$,and
$p_2$, $m_2$ are the momenta and the masses of the
decay quark, the final one and the spectator
respectively; $\Gamma_1^{\mu}$ is the weak interaction vertex and to the
lowest order, $\Gamma_1^{\mu}$ has the form of
$\gamma_{\mu}(1-\gamma_5)$ for the charged current.

As pointed out above, the B-S wave function $\chi_p(q)$ of the heavy
quark pair system under the instantaneous
approximation, can be evaluated by solving the corresponding
Schr\"odinger equation with a QCD inspired potential.
Usually, to make the nonrelativistic instantaneous
approximation for the B-S equation is implied in the
rest frame of the concerned bound state.

Now let us
outline the the approximation but in a form, which
may be used further.

The instantaneous approximation is to carry through
the integration over the $q_0$ component for the B-S equation Eq.(2)
firstly,
when the kernel
at the rest frame has a simple form
\begin{equation}
V(p,k,q) \sim V(|{\bf k}-{\bf q}|),
\vspace{5mm}\end{equation}
then as a result, the B-S equation Eq.(2)
is deduced into a three-dimensional
equation, i.e., the Schr\"odinger equation in a
momentum representation$^{[14]}$.

However, as pointed out above, when the decays with sizable recoil
are concerned, it still is not sufficient to have the relation
between the wave functions in the rest frame only, because in
the Mandelstam formalism Eq.(5), the start point of our approach, at least
one of the B-S wave functions in a Lorentz covariant form in
a moving frame is required.
To pursue the purpose, we take the way to make the instantaneous
approximation to the whole matrix element
itself instead of only the wave functions.

As the first step, we need to divide the relative
momentum $q$ of the B-S equation
into two covariant parts firstly: $q_{\pa}$ and $q_{\pe}$,
a parallel one and an orthogonal one to the centre mass momentum
$p$ respectively, i.e.,
\begin{equation}
\begin{array}{l}
q^\mu =q^\mu _{\pa} +q^\mu _{\pe}
\end{array}
\vspace{2mm}\end{equation}
where $q^\mu _{\pa}\equiv \dfr{p\cdot q}{M^2_p}p^\mu $;
$q^\mu _{\pe}\equiv q^\mu -q^\mu _{\pa}$.
Correspondingly, we have two Lorentz invariant variables:
\begin{equation}
\begin{array}{l}
q_p=\dfr{p\cdot q}{M_p},\\[0.2in]
q_{\pet}= \sqrt{q_p^2-q^2}=\sqrt{-q^2_{\pe}}.
\end{array}
\vspace{2mm}\end{equation}
In the rest frame of the meson, i.e. ${\bf p}=0$, they turn back
to the usual component $q_0$ and
$|{\bf q}|$ respectively.

Now in terms of these
variables, the covariant form of the
wave function can be obtained.
The volume element of the relative momentum $k$
can be written in an invariant form:
\begin{equation}
d^4k=dk_pk_{\pet}^2dk_{\pet}dsd\phi,
\vspace{2mm}\end{equation}
where $\phi$ is the azimuthal angle,
$s=\dfr{k_pq_p-k\cdot q}{k_{\pet}q_{\pet}}$.
The interaction kernel can be denoted as:
\begin{equation}
V(|{\bf k}-{\bf q}|)  =V(k_{\pe},s,q_{\pe}),
\vspace{2mm}\end{equation}
which is independent of  $k_p$ and $q_p$.
Let us introduce the notation for convenience:
\begin{equation}
\begin{array}{lcl}
\varphi_p({ q}^\mu _{\pe}) & \equiv & i \dsp\int \dfr{dq_p}{2\pi}
\chi_p(q^\mu _{\pa}, { q}^\mu _{\pe}),\\[0.3in]
\eta({ q}^\mu _{\pe}) & \equiv & \dsp\int \dfr{k^2_{\pet}dk_{\pet}ds }
{(2\pi)^3}
  V(k_{\pe},s,q_{\pe}) \varphi_p({ k}^\mu _{\pe}) ,  \\[0.2in]
\end{array}
\vspace{3mm}\end{equation}
thus the B-S equation can be rewritten in short as:
\begin{equation}
\chi_p(q_{\pa},q_{\pe})=S_1(p_1)\eta(q_{\pe})S_2(p_2),
\vspace{2mm}\end{equation}
where $S_1(p_1)$ and  $S_2(p_2)$ are the propagators of the
constituent particles
and they can be decomposed as:
\begin{equation}
S_i(p_i)=\dfr{\Lambda^{+}_{ip }( q_{\pe}) }{ J(i)  q_p+
\alpha_iM-\omega_{ip}+i\epsilon}
+ \dfr{\Lambda^{-}_{ip}( q_{\pe}) }{ J(i)   q_p+
\alpha_iM+\omega_{ip}-i\epsilon},
\vspace{2mm}\end{equation}
with
\begin{equation}
\begin{array}{l}
\omega_{ip} = \sqrt{m_i^2+q_{\pet}^2}\;,\\[0.2in]
\Lambda^{\pm}_{ip}( q_{\pe} ) = \dfr{1}{2\omega_{ip} } \left[
\dfr{\rlap/{p} }{ M}\omega_{ip} \pm J(i) (m_i+ \rlap/{q_{\pe}} ) \right],
\end{array}
\vspace{2mm}\end{equation}
$i=1,2$ and $J(i)=(-1)^{i+1}$

Here $\Lambda^{\pm}_{ip}( q_{\pe} )$ satisfies the following relations:
\begin{equation}
\begin{array}{l}
\Lambda^{+}_{ip}(q_{\pe} ) + \Lambda^{-}_{ip}(q_{\pe} )
 =\dfr{\rlap/{p} }{M},\\[0.2in]
\Lambda^{\pm}_{ip}(q_{\pe} ) \dfr{\rlap/{p} }{M}
\Lambda^{\pm}_{ip}(q_{\pe} ) = \Lambda^{\pm}_{ip} (q_{\pe} ),\\[0.2in]
\Lambda^{\pm}_{ip}(q_{\pe} ) \dfr{\rlap/{p} }{M}
\Lambda^{\mp}_{ip}(q_{\pe} ) = 0.
\end{array}
\vspace{2mm}
\end{equation}
Thus, $\Lambda^{\pm}_{ip}( q_{\pe} )$ may be referred to as $p$-projection
operators ( $p$ is the momentum
of the bound state )
while in the rest frame they correspond to the energy projection
operators.

If defining $\varphi^{\pm\pm}_{p}(q_{\pe})$ as
\begin{equation}
\varphi^{\pm\pm}_{p}(q_{\pe}) \equiv
\Lambda^{\pm}_{1p}(q_{\pe} ) \dfr{\rlap/{p} }{M}
\varphi_{p}(q_{\pe})
\Lambda^{\pm c}_{2p}(q_{\pe} ) \dfr{\rlap/{p} }{M},
\vspace{2mm}\end{equation}
where the upper index $c$ denotes the charge conjugation,
under the notation:
$$\Lambda^{\pm c}_{2p}(q_{\pe} ) \equiv  \Lambda^{\pm }_{2p}(q_{\pe} ),$$
and integrating over $q_p$ on both sides
of Eq.(12), we obtain
\begin{equation}
\begin{array}{lcl}
(M-\omega_{1p}-\omega_{2p}) \varphi^{++}_{p}(q_{\pe}) &=&
\Lambda^{+}_{1p}(q_{\pe} )
\eta_{p}(q_{\pe}) \Lambda^{+c}_{2p}(q_{\pe}), \\[0.2in]
(M+\omega_{1p}+\omega_{2p}) \varphi^{--}_{p}(q_{\pe}) &=&
\Lambda^{-}_{1p}(q_{\pe} )
\eta_{p}(q_{\pe}) \Lambda^{-c}_{2p}(q_{\pe}), \\[0.2in]
\varphi^{+-}_{p}(q_{\pe})=\varphi^{-+}_{p}(q_{\pe})=0. & &
\end{array}
\vspace{2mm}\end{equation}
The normalization condition becomes into the following covariant form:
\begin{equation}
\int\dfr{q^2_{\pet}dq_{\pet} }{ 2\pi^2}{\rm tr} \left[
\overline{\varphi}^{++}_{p}(q_{\pe}) \dfr{\rlap/{p} }{M}
\varphi^{++}_{p}(q_{\pe}) \dfr{\rlap/{p} }{M} -
\overline{\varphi}^{--}_{p}(q_{\pe}) \dfr{\rlap/{p} }{M}
\varphi^{--}_{p}(q_{\pe}) \dfr{\rlap/{p} }{M} \right ] =2M.
\vspace{2mm}\end{equation}

Now for the usage later on,
let us introduce two 3-momenta $\tilde{\rm\bf p}_1$ and
$\tilde{\rm\bf p}_2$,
\begin{equation}
\begin{array}{l}
\tilde{\rm\bf p}_1 \equiv
\dfr{\omega_{1p} }{M}{\rm\bf p} + {\rm\bf q}_{\pe},\\[0.2in]
\tilde{\rm\bf p}_2 \equiv
\dfr{\omega_{2p} }{M}{\rm\bf p} - {\rm\bf q}_{\pe}.
\end{array}
\vspace{2mm}\end{equation}
In the case of the weak binding as concerned here,
the wave functions of
the double heavy quark systems
can be constructed approximately as follows:
\begin{equation}
\begin{array}{lcl}
\varphi^{\lambda ++}_p(q_{\pe}) &=& \dsp\sum_{ss'}
\dfr{1 }{\sqrt{4\omega_{1p}\omega_{2p}} }u_s(\tilde{\rm\bf p}_1)
\bar{v}_{s'}(\tilde{\rm\bf p}_2)
\phi_p^+(q_{\pet})\chi^{\lambda}_{ss'},\\[0.4in]
\varphi^{\lambda --}_p(q_{\pe}) &=& \dsp\sum_{ss'}
\dfr{1 }{\sqrt{4\omega_{1p}\omega_{2p}} }v_s(\tilde{\rm\bf p}_1)
\bar{u}_{s'}(\tilde{\rm\bf p}_2)
\phi_p^-(q_{\pet})\chi^{\lambda}_{ss'},\\[0.2in]
\end{array}
\vspace{2mm}\end{equation}
where $u_s(\tilde{\rm\bf p}_i)$,
${v}_{s'}(\tilde{\rm\bf p}_i)$ (i=1,2) are the Dirac spinors
of free particles with masses $m_i$ and momenta
$\tilde{\rm\bf p}_i$; $\chi^{\lambda}_{ss'}$ is the
Clebsch-Gordan coefficients
that make $s'$ and $s$ couple to $\lambda$;
and $\phi^{\pm}(q_{\pet})$ is the scalar
part of the wave function.

In the present case, owing to weak binding,
the $\varphi^{\lambda --}_p(q_{\pe})$
is a small component and can be ignored. In fact, if the kernel is of
scalar and/or vector, the $\varphi^{\lambda --}_p(q_{\pe})$
is in the order of $(v/c)^4$ to $\varphi^{\lambda ++}_p(q_{\pe})^{[18]}$.
Furthermore, if ignoring the components
proportional to the $q_{\pe}$ in the spinor structure
due to the nonrelativistic nature,
$\varphi^{\lambda ++}_p(q_{\pe})$ can be simplified:
\begin{equation}
\varphi^{\lambda ++}_p(q_{\pe}) =\dfr{\rlap/{p}+M }{ 2\sqrt{2} M}
(\alpha\gamma_5 +\beta \rlap/{\epsilon} )\phi(q_{\pet})
\vspace{2mm}\end{equation}
where $\alpha=1,\;\beta=0$ for an $^1S_0$ state and
$\alpha=0,\;\beta=1$ for a $^3S_1$ state,
while the ``radius'' wave
function $\phi(q_{\pet})$ satisfies the following
Schr\"{o}dinger equation:
\begin{equation}
\dfr{q^2_{\pet}}{2\mu}\phi(q_{\pet}) +\int \frac{k^2_{\pet}dk_{\pet}ds }{
(2\pi)^3 } V(s,k_{\pe},q_{\pe}) \phi(k_{\pet}) =E\phi(q_{\pet}),
\vspace{2mm}\end{equation}
with the reduced mass $\mu=\dfr{m_1m_2}{m_1+m_2}$ of the system.

Thus, we have established the relation in covariant form
between the B-S wave function and the solution of the
Schr\"odinger equation
with the instantaneous approximation.

Now we start the procedure to make the instantaneous approximation
to the whole matrix element itself.
The weak current
matrix element,
by integrating over the $q_0$ component of Eq.(10)
with the method of contour integration.

As asserted above, the negative energy parts of the wave functions
are very small in weak binding cases so that we can ignore their
contributions in the lowest order approximation.
In the present case, we are concerning weak-binding bound states only,
thus those contributions from the negative energy parts will be ignored.
Putting Eqs.(12), (13) into Eq.(5), we have the matrix element:
\begin{equation}
\begin{array}{lcl}
l_{\mu}(r)&=&\dsp\int \dfr{d ^4 q}{(2\pi)^4} \left[\bar{\eta}'_{p'}(q'_{\pee})
 \dfr{\Lambda^{+}_{1p'}( q'_{\pee}) }{  q'_{p'}+
\alpha_1'M'-\omega_{1'p'}+i\epsilon} \Gamma_{1\mu} \right.\\[4mm]
 & & \left. \dfr{\Lambda^{+}_{1p}( q_{\pe}) }{  q_p+
\alpha_1 M+\omega_{1p}+i\epsilon}\cdot
 \dfr{\Lambda^{+}_{2p}( -q_{\pe}) }{  q_p-
\alpha_2 M+\omega_{2p}-i\epsilon}\right]{\eta}_{p}(q_{\pe}).
\end{array}
\vspace{3mm}\end{equation}

In the bracket of the integrand, there are three poles
at points $a_i$ in the complex-$q_0$ plane:
\begin{equation}
\begin{array}{l}
a_1=-\alpha_1 M +\omega_1 -i \epsilon,\\[4mm]
a_2=\alpha_2 M -\omega_2 +i \epsilon,\\[4mm]
a_1'=\alpha_2 M -E' \sqrt{({\rm\bf r} + {\rm\bf q})^2 +m_1^{'2} } -
i\epsilon,\\[4mm]
\end{array}
\vspace{2mm}
\end{equation}
and two branch cuts starting at the branch points:
\begin{equation}
q_0 \simeq m_2\pm i \dfr{m_1'}{\gamma},
\vspace{2mm}
\end{equation}
with $\gamma\equiv \dfr{|{\rm\bf r}|}{M'}$
and all the terms relevant to ${\rm\bf q}$ are ignored, due to the fact
$\omega'_{1p'}=\sqrt{q_{\pee}^{'2} +m_1^{'2} }$.

Now the problem becomes how to carry
out the integration over the $q_0$ component
on the right side of Eq.(23) or, in another words,
how to treat the cuts of the integrand if contour integration
method is adopted.
Wereas here we will treat the branch cuts in the integrand
approximately, by expanding $\omega'_{1p'}$ as follows:
\begin{equation}
\omega'_{1p'}= \sqrt{q_{\pee}^{'2} +m_1^{'2} }=m_1'+
\dfr{q_{\pee}^{'2}}{2m_1'} +\cdots\;\;.
\vspace{2mm}
\end{equation}
In a weak binding limit it is quite good approximation.
According to Cauchy's theorem,
the integration of a closing contour on the upper half plane
of the complex-$q_0$ for the matrix element $l_\mu(r)$,
is just that summing up
all poles' residues.
However, as the pole $a_2$ on the upper half plane is very close to
the pole $a_1$ on the lower half
plane, i.e. the distance
\begin{equation}
\Delta\equiv a_1-a_2\simeq M-m1-m2 +\dfr{\rm\bf q^2}{2\mu_1}
\vspace{4mm}\end{equation}
is small,
the value of the integration is dominated by the residue of the pole
$a_2$ only. The contribution from the pole $a'$ is not important
i.e. may be ignored at all at the concerned accurate level.\footnote{
In fact, some poles ( even cuts )
may be induced into the matrix element through $\bar{\eta}'_{p'}(q'_{\pee})$,
however as the similar reason as here, they would not contribute
so substantially as that of the pole $a_2$
to the final results in all the cases of weak binding. }.
Therefore, we obtain:
\begin{equation}
l_{\mu}(r)=\dsp\int \dfr{d^3{\rm\bf q}}{(2\pi)^3}
\bar{\eta}'_{p'}(q'_{\pee})
\dfr{
\Lambda^{+}_{1p'}( q'_{\pee}) \Gamma_{1\mu}
 \Lambda^{c+}_{2p}(-q_{\pe})
 \Lambda^{+}_{1p}( q_{\pe}) }{
 ( q'_{p'}+ \alpha_1'M'-\omega'_{1p'}) (M-\omega_{1p}-\omega_{2p})}
{\eta}_{p}(q_{\pe})\; .
\vspace{2mm}
\end{equation}
It is easy to prove the following relations
\begin{equation}
\begin{array}{rcl}
  q'_{p'}+ \alpha_1'M'-\omega'_{1p'}&=& M'-\omega'_{1p'}-\omega'_{2p'},\\[4mm]
  \Lambda^{c+}_{2p}(-q_{\pe})
&=&
\dfr{\omega'_{2p'}}{\omega_{2p} }
  \Lambda^{c+}_{2p'}(-q'_{\pee})\gamma_0 \Lambda^{c+}_{2p}( -q_{\pe})\;,
\end{array}
\vspace{2mm}
\end{equation}
and with Eq.(17) from Eq.(28)
we obtain
the required equation:
\begin{equation}
l_{\mu}(r)\approx \int \dfr{q^2_{\pet} dq_{\pet} ds}{(2 \pi)^2}
 \dsp {\rm tr} \;[\;\overline{\varphi}^{++}_{p'}(q'_{\pee} ) \Gamma_{\mu}
{\varphi}^{++}_{p}(q_{\pe}) \dfr{\rlap/{p}}{M} \;]\;
\dfr{\omega'_{2p'} }{\omega_{2p} },
\vspace{2mm}\end{equation}
where $q'_{\pee}$, $\omega'_{1p'}$, $\omega'_{2p'}$,
$\omega_{1p}$, $\omega_{2p}$
are as expressed in Eqs.(8), (14) and (26) i.e.
\begin{equation}
\begin{array}{l}
\omega'_{1p'}=\sqrt{q_{\pee}^{'2} +m_1^{'2}},\\[0.2in]
\omega_{ip} =\sqrt{{\rm\bf q}^2+m_i^2},\\[0.2in]
\omega'_{2p'} =\dfr{E'\omega_{2p}+{\rm\bf {r\cdot q} }}{M'}, \\[0.2in]
q'_{\pete}=\sqrt{\omega^{'2}_{2p'}-m_2^2}.
\end{array}
\vspace{2mm}\end{equation}

We should note here that based on
the adopted extra approximations such as Eq.(25), the
Eq.(30) is valid only with not too large recoils, i.e.
$\gamma\equiv \dfr{|\bf\rm r|}{M'} \leq 1$,
however most of the decays of double heavy quark systems
such as our concerning
processes $B_c\to J/\psi+X$,
$B_c\to B_s +X$ etc. satisfy the condition well.\footnote{Here in order
to show the approach emphasized in this paper briefly so as to see
the inspire of the approach clearly we take the extra approximation,
as matter of fact it may be weaken or improved
in the practical usage.}
Using the wave functions in the form of Eq.(20) for both of the initial
state and the final state, it follows that
\begin{equation}
l_{\mu}(r)=\int \dfr{q^2_{\pet} dq_{\pet} ds}{(2 \pi)^2}
\;[\;\overline{u}_l (\tilde{\rm\bf p'}_1)  \Gamma_{\mu}
{u}_m (\tilde{\rm\bf p}_1) \;]\; \overline{\varphi}^{+}_{p'}(q'_{\pete} )
{\varphi}^{+}_{p}(q_{\pet})
\chi^\lambda_{ls'} \chi^{\lambda'}_{s'm}
(\dfr{\omega'_{2p'} }{4\omega_{2p} \omega_{p} \omega'_{1p'} }  )^{\frac{1}{2}}
\vspace{2mm}\end{equation}
where
\begin{equation}
\begin{array}{l}
\omega_{1p'} =\sqrt{q^{'2}_{\pete}+m^{'2}_1}\\[0.2in]
p_1=(\omega_1, {\rm\bf q})\\[0.2in]
p'_1=\dfr{\omega'_{1p}+\omega'_{2p} }{M'}p'-\dfr{\omega_{1p}+ \omega_{2p} }
{M}p+p_1,
\end{array}
\vspace{2mm}\end{equation}
and the normalization condition for the `spectator' ( the antifermion
with momentum $p_2$ in Fig. 1 ),
\begin{equation}
\bar{v}_{s'}(p_2)\dfr{\rlap/p}{M} v_{l'}(p_2)=2\omega_2 \delta_{s'l'},
\vspace{2mm}\end{equation}
has been used.

In order to compare with those results of Refs.[16,17],
we compute out the form factors and it is very interesting that
we may extract a similar `universal'
Isgur-Wise function at last.

There is some arbitrariness in choosing the directions of the spins
of the quarks when calculating the form factors. Thus we may use it to
simplify the calculations.
It is convenient to take direction of the spins orthogonal to the
$p$ and $p'$ because the spins in this direction remains unchanged if
a Lorentz boost along the $p'$ direction is taken.

In the Appendix of
Ref.[7], a covariant formalism to calculate the creation of a pair of
fermion-antifermion have been derived in the spirit of helicity
amplitude$^{[19]}$. We will employ the method for the present problem.

A similar formalism can be obtained for a fermion scattered by a virtue
$W^{\pm}$ boson i.e.
the amplitudes with  possible spin directions read as
\begin{equation}
\begin{array}{l}
M^{\mu}_{1,2}=L_{+}\dfr{1}{2}{\rm tr}\left[ (\rlap/{p'_1}+m'_1)
\dfr{1\pm \gamma_5\rlap/{k_1} }{2} (\rlap/{p_1}+m_1)
\Gamma^{\mu}\right]\\[0.2in]
M^{\mu}_{3,4}=L_{-}\dfr{1}{2}{\rm tr}\left[ (\rlap/{p'_1}+m'_1)
\gamma_5\dfr{1\pm \gamma_5\rlap/{k_1} }{2} (\rlap/{p_1}+m_1)
\Gamma^{\mu}\right]
\end{array}
\vspace{2mm}\end{equation}
where
\begin{equation}
\L_{\pm}=[\dfr{1}{2}(p_1\cdot p'_1 \pm m_1m'_1)]^{-\frac{1}{2}},
\vspace{2mm}\end{equation}
and $k_1$ is an auxiliary and space-like vector ($k_1^2=-1$),
which is orthogonal to the initial
momentum $p_1$ and the final one $p'_1$.
It is easy to see that the first equation of Eq.(25) describes
the spin nonflip amplitude while the second
describes a flip one.
Both fermions are fully polarized along the $\pm k_1$ directions.
It should be noted here
that in these formulae the relative phases of the spinors
among those states with different polarizations have been fixed.

To calculate the form factors from a current matrix element,
we need to
construct the spin wave functions of the individual quarks into
definite spin ones to describe
the initial and the final states of the matrix element respectively.
In general, for an $^1S_0$ state
\begin{equation}
\chi_{ss'}=\dfr{1}{\sqrt{2} }( \uparrow\downarrow -\downarrow\uparrow ),
\vspace{2mm}
\end{equation}
and for an $^3S_1$ state, the spin structure corresponding to the
three possible independent polarizations are
\begin{equation}
\begin{array}{l}
\chi^{k_1}_{ss'}=\dfr{1}{\sqrt{2} }( \uparrow\downarrow +\downarrow\uparrow ),
\\[0.2in]
\chi^{k_2}_{ss'}=\dfr{1}{\sqrt{2} }( \uparrow\uparrow +\downarrow\downarrow),
\\[0.2in]
\chi^{k_3}_{ss'}=\dfr{i}{\sqrt{2} }( \uparrow\uparrow -\downarrow\downarrow),
\end{array}
\vspace{2mm}
\end{equation}
where three vectors $k_1$, $k_2$, and $k_3$ orthogonal to each other,
are used to denote the polarization directions of the $^3S_1$ state.

Thus, for a transition $P\rightarrow P' +X$, the amplitude reads
\begin{equation}
M^{\mu}_0 =L_+ \dfr{1}{2} {\rm tr} [\;(\rlap/p'_1+m'_1) \gamma_5 \gamma_5
(\rlap/p_1+m_1)\Gamma^{\mu}\;],
\vspace{2mm}\end{equation}
and, for a transition $P\rightarrow V +X$, the corresponding amplitudes read
\begin{equation}
\begin{array}{l}
M^{\mu}_1 =L_+ \dfr{1}{2} {\rm tr} [\;(\rlap/p'_1+m'_1) \gamma_5 \rlap/k_1
(\rlap/p_1+m_1)\Gamma^{\mu}\;],\\[0.2in]
M^{\mu}_2 =L_- \dfr{1}{2} {\rm tr} [\;(\rlap/p'_1+m'_1) \rlap/k_1
(\rlap/p_1+m_1)\Gamma^{\mu}\;],\\[0.2in]
M^{\mu}_3 =L_- \dfr{1}{2} {\rm tr} [\;(\rlap/p'_1+m'_1) \gamma_5
(\rlap/p_1+m_1)\Gamma^{\mu}\;],
\end{array}
\vspace{2mm}\end{equation}
where $M_1$, $M_2$ and $M_3$ correspond to those of various
projections of the polarizations
of the final states. In fact, if letting
\begin{equation}
\begin{array}{l}
k_2^{\mu}=\dfr{1}{2}L_+L_-\epsilon^{\mu\nu\rho\sigma}p'_{1
\nu} k_{1\rho} p_{\sigma},\\[0.2in]
k_3^{\mu}=\dfr{L_+L_-}{2M'}[ (p'\cdot p'_1)  p_1^{\mu} -(p'\cdot p)
p^{'\mu}_1 ]
\end{array}
\vspace{2mm}\end{equation}
correspond to the polarizations of $M_2$ and $M_3$,
the amplitudes of the $P\to V+X$
transition can be written down in a compact form:
\begin{equation}
M^{\mu} =L_+ \dfr{1}{2} {\rm tr} [\;(\rlap/p'_1+m'_1) \gamma_5 \rlap/\epsilon'
(\rlap/p_1+m_1)\Gamma^{\mu}\;],
\vspace{2mm}\end{equation}
where
\begin{equation}
\begin{array}{l}
\epsilon^{'\mu}=\epsilon^{\mu} - \dfr{C \dot (\epsilon \cdot p )}
{p^2-\dfr{(p\cdot p')}{m^{'2} } }p^{\mu}_{\pee}  ,\\[0.4in]
C=\dfr{L_+} {\sqrt{\dfr{(p'\cdot p_1)(p' \cdot p'_1)}{M^{'2}}-\dfr{1}{L_-^2}}}
-1,\\[0.4in]
p_{\pee}=p -\dfr{(p\cdot p')}{M^{'2} } p'.
\end{array}
\vspace{2mm}\end{equation}

After a straightforward calculation,
the form factors may be formulated:
\begin{equation}
\begin{array}{lcl}
f_{\pm}&=&\xi \left[ \dfr{1}{M}(1-\dfr{m_2}{m'_1})\pm \dfr{\omega'_1+\omega'_2}
{M'm'_1} \right],\\[0.4in]
g&=&\xi \dfr{\omega'_1+\omega'_2}
{M M'm'_1} ,\\[0.4in]
f&=& \xi \left( \dfr{(p\cdot p'_1)}{M m'_1} +1\right),\\[0.4in]
a_{\pm}&=&\xi \left[ \dfr{2m_2}{M^2m'_1}+ \delta \mp
\left(\dfr{\omega'_1+\omega'_2} {MM'm'_1} +\dfr{(p\cdot p')}{M^{'2}}\delta
\right) \right],
\end{array}
\vspace{2mm}\end{equation}
where
\begin{equation}
\delta=-\dfr{C \dot ( 1 + \dfr{(p\cdot p'_1)}{Mm'_1} ) }{ p^2 -
\dfr{(p\cdot p')^2}{M^{'2}} }.\\[2mm]
\end{equation}
In the case of the zero recoil vicinity
(${\rm\bf r}\to 0$)
\begin{equation}
\delta\to -\dfr{m_2}{M^2m'_1},
\vspace{2mm}
\end{equation}
while the ``common'' factor, which may be considered as the
`universal' Isgur-Wise function, is written in the frame of the initial
meson at rest ( ${\rm\bf p}=0$ ):
\begin{equation}
\xi=\left( \dfr{2\omega'_2m^2_1m^{'2}_1 }{((p_1\cdot p_1')+m_1m'_1)
\omega_1\omega'_1\omega_2} \right)^{\frac{1}{2} }
\int\dfr{d^3{\rm\bf q}}{(2\pi)^3}
\phi^{'\ast}_{p'}(q'_{\pete})\cdot \phi_p (|{\rm\bf q}|),
\vspace{3mm}\end{equation}
where $\phi'^{\ast}_{p'}(q'_{\pete})$ and $\phi_p (|{\rm\bf q}|)$
correspond to the radius parts of the
wave functions of the meson in
the initial state and the one in the final state, respectively.
Being in a covariant form and in the C.M.S. one, they
may be obtained by
solving the Schr\"{o}dinger
equation Eq.(22) and its specific one in its center mass system
respectively, as long as the QCD inspired potential is transformed into
the corresponding form in the equation. We should
note i). Of the approximations, the instantaneous one
on the whole matrix element, is essential in the
approach, and we think it is relible to the purposes
due to the facts that we restrict to apply this approach only
to the weak binding systems and the approximation only for the B-S equation,
is proven to work well in many cases as long as a weak binding
system is concerned$^{[14]}$.
ii). The wave functions obtained by the present way
are more reliable than those by other ways,
because the adopted potential is proven to work
well for heavy quark systems,
although the `universal' function $\xi$ being the
overlap integration of the wave functions, does not
sensitive to the specific radius wave functions very much, especially
when the momentum recoil is not great it is contralled by the
normalization condition of the wave functions,
i.e. the overlap integration is approaching to
the normalization when
the recoil is approaching to zero and
the reduced mass of the final state is approaching
to that of the initial one in the meantime.
iii). When carrying through the trace for
$\gamma$ matrices in the Eqs.(39) and (42) so as to reach to the form
factos, the contributions from $(\vec{\gamma}\cdot {\rm \bf q})$ in
the terms $\rlap/{p_1'}$ and $\rlap/{p_1}$,
have been ignored savely in the considered
accuracy, because they are small in the case of
weak binding, and when carrying through the
integration Eq.(47), of the integrand all terms
proportional to ${\rm \bf q}$ in
odd power will not contribute at all
( only even power terms contribute ).

The proposed approach
to calculate the weak decay matrix elements, so as to
the form factors is completed. It is expected that the
approach is available as long as
the mesons in initial and final states are of weak binding.
It is interesting to compare ours with
that of IGSW model$^{[17]}$.
In their model, the authors calculated the form factors by using the
Gaussian-type
wave functions, of which the parameters
are determined by the variational method.
It is easy to see that
in the case of weak binding and at zero recoil vicinity,
the formalism of our approach is consistent
with theirs except a tiny difference in
the formulation for the overlap integration of the
wave functions$^{[17][20]}$.
However still in the case of
weak binding but with a large recoil, there are two remarkable
deviations between the
two approaches. One deviation
comes from the difference in the spin structure
of the wave functions. The other
from the arguments in the wave function integrand. For instance,
the function
corresponding to the $\xi$ in IGSW model is $\sqrt{\dfr{M_{B_c}}{M'}} F_3$
and reads$^{[17]}$:
\begin{equation}
\xi_{IGSW}=\left( \dfr{2\beta\beta'}{\beta^2+\beta^{'2} }\right)^\frac{3}{2}
\exp^{ -\dfr{m_2^2}{2{\rm \tilde{M}\tilde{M}'
}}\dfr{t_m-t}{\kappa^2(\beta^2+\beta^{'2})} }\;\;,
\vspace{2mm}\end{equation}
where $t_m=(M-M')^2$; $\rm \tilde{M}$ and  ${\rm \tilde{M}'}$
are the masses of the ``mock meson''$^{[17]}$;
$\beta$ and $\beta'$ are the variational parameters for the initial
and the final states
respectively; $m_2$ is the mass of the `spectator' and
$\kappa$ is a parameter introduced by hand.
In the
IGSW model, $\kappa$ is adjusted to be 0.7 by fitting the $\pi$
electromagnetism form factor and the authors
of Ref.[17] regarded it as relativistic corrections due to a large
recoil. However in our approach it is different
i.e. all factors come into the formula automatically.
The interesting thing is that occasionally
the numerical calculation shows that the function of $\xi$ obtained
in our approach, is
very close to that of the IGSW model with $\kappa=0.7$.
To show the fact, we present the corresponding $\xi$ functions of
$B_c\to J/\psi+X$ and $B_c\to B_s+X$ in Fig.2 and Fig.3 respectively.
The dotted line represents the $\xi$ function obtained by
the Gaussian wave
functions and with the original formalism of Ref.[17]
and $\kappa=1.$ as well. The dashed line represents
that obtained by the wave functions
from solving the Schr\"odinger equation of the potential model$^{[21]}$
with the same formalism. The dotted-dashed line
represents that of the IGSW model but with
$\kappa=0.7$. Note that the parameters $\beta_{B_c}=0.88$ and
$\beta_{J/\Psi}=0.65$ which were obtained
by the authors of Ref.[22] are used
here for the dotted, dashed and dotted-dashed three curves.
The solid line represents the $\xi$ function achieved by
our approach with the wave functions obtained by the potential model$^{[23]}$.
It is easy to see from the Fig.2 that
the result of Eq.(47), is very close to that of
IGSW model with
$\kappa=0.7$. It means that our results involve
reasonable effects automatically.
In order to have more comparison we also show
the $\xi$ function
for $B_c\to B_s +X$ in Fig.3 and the meaning of each line is the
same as in Fig.2, although we do not expect that
our approach is so suitable as that for the formal decays $B_c\to J/\psi+X$,
due to the s quark is not so heavy.

It is also very interesting to
see the behavior in the limit when the quark mass is approaching
to infinite because it will let us see the relation between
the form factors obtained here and the
universal Isgur-Wise function. At
the limit of
\begin{equation}
m_2<< m_1,m_1' \;\;\;\;\;{\rm and}\;\;\;\;m_1,m_1'\to \infty,
\vspace{2mm}\end{equation}
the formulae of Eqs. (44) -- (47) reproduce those of the Isgur-Wise
formalism$^{[1-6]}$ for the form factors.
In fact, in the limitation of Eq.(49) and
from Eq.(33) and Eq.(43), we have
\begin{equation}
\begin{array}{l}
p_1\to m_1\cdot v,\\[0.2in]
p_1'\to m_1'\cdot v',\\[0.2in]
\epsilon' \to \epsilon'.
\end{array}
\vspace{2mm}\end{equation}
Hence the Eq.(42) can be rewritten as
\begin{equation}
l_{\mu}(r)=\xi(v\cdot v'){\rm tr}\left[ (1+\rlap/{v'})
(\alpha +\beta \gamma_5\rlap/{\epsilon}) (1+\rlap/{v}) \right],
\vspace{2mm}\end{equation}
where $\xi(v\cdot v')$ is just the universal Isgur-Wise function and
has the following form:
\begin{equation}
\xi(v\cdot v')=\dfr{\sqrt{2 v\cdot v'} }{\sqrt{1+v\cdot v'} }
\dsp\int \dfr{d^3{\rm\bf q}}{(2 \pi)^3} \phi'^\ast(q'_{\pete})
\phi (|{\bf\rm q}|).
\vspace{2mm}\end{equation}
Thus the Eq.(51) and Eq.(52)
reproduce those of the Isgur-Wise's formalism
in the infinite heavy limit$^{[1-3]}$,
so a link between the Isgur-Wise function and the nonrelativistic
wave function overlap integration has been established. A factor of
$[0.5({1+v\cdot v'} )]^{-\frac{1}{2}}$
has been derived by Bjorken$^{[24]}$ by using the Cabibbo-Radicati
sum rule$^{[25]}$, however
Here in our formalism it automatically appears in Eq.(52).

In Ref.[9], we have
applied the approach to calculate the weak decays
of the $B_c$ systematically.
It is no doubt that when the experimental
study of $B_c$ meson has fruitful results, our approach will receive
serious tests.
We would emphasize here that for the
decay modes $B_c\rightarrow J/\Psi(\eta_c) +X$, which might
be used as the most typical channels to
identify the $B_c$ meson later on
in experiments, the results from our approach
are the most reliable ones,
as in the calculation of the form factors
the adopted nonrelativistic wave functions obtained from the
potential model are tested well in addition to our approach
itself is suitable them.
Although our predictions to the processes are very
close to those of IGSW model with $\kappa=0.7^{[20]}$,
there exists some deviations in numerical results
for the processes $B_c\to B_s +X$,
i,e, our predictions to the laters
are larger than those of Ref.[20] both for semileptonic decays
and nonleptonic decays. The deviations
may be understood as follows: although the $\xi$ function for the
modes is smaller than that of the IGSW model
as shown in Fig.3, the form factors
gain an enhancement from the spinor factor as shown
in Eq.(39) and Eq.(42).

We should point out that the approach has quite
wide potential applications, as long as
the binding is weak . For example,
the approach may be applied to
those decays between the baryons which contain two heavy
quarks in addition to the $B_c$ meson's by means of
a heavy diquark picture,
and we will describe it
in Ref.[26].

As stated above, when taking the instantaneous approximation
to the whole matrix element by doing the integration of $q_0$ on
Eq.(23), an extra approximation has been adopted i.e. to take the
expansion Eq.(26), thus the branch points are alternated.
The extra approximation
set down one more restraint
$$\dfr{q_{\pee}^{'2}}{m_1^{'2}} \ll O(1)  $$
i.e. the momentum recoil cannot be
too great. If treating the branch points in the integrand by an
accurate method i.e. making cuts, in principle, we may integrate out
$q_0$ exactly. However, the potential models only offer us the numerical
solutions owing to the complicated concrete potential, thus we
have to deal with an integration not analytically throughout,
thus we cannot guarantee results so accurate by numerically
integrating a counter with cuts, finally as a result the accurateness
is lost too. We would conclude that the extra approximation
may be weaken, but not be avoid at all$^{[26]}$. The experimental
information on various decays is desired very much now, as we are
at the position, that
the approach remain to be tested and improved guiding
by the information. In the near future the experiments
especially on the $B_c$ decays are very important.
Having the information, we will learn much on the decay
mechanism as will as on
the approach.

\bigskip
\vspace{6.0ex}
{\Large\bf{ Acknowledgements}\vspace{1ex}}

The authors would like to thank
Y.-B. Dai and K.-T. Chao for helpful discussions.

\newpage
\centerline{\large\bf{ Figure Captions}}
\vspace{1ex}

\begin{enumerate}

\item  Feynman diagram corresponding to the weak current matrix element
sandwiched by the $B_c$ meson state as the initial state and a single
particle state of the concerned final state.

\item  The $\xi$ function for $B_c\to J/\psi+X$. The dashed line:
the $\xi$ function of IGSW model with $\kappa=1.$; the
dotted-dashed line: the $\xi$ function obtained by the
formalism of Ref.[18] with
the wave function solved by potential I$^{[12]}$;
the doted line: IGSW  model with
$\kappa=0.7$;  the solid line: the $\xi$ function obtained by our approach.

\item  The $\xi$ function for $B_c\to B_s + X$.
The meaning of each type line is the same as that in Fig. 3.

\end{enumerate}

\vspace{8ex}

\newpage
 \begin{thebibliography}{7}

\bibitem{1}  N. Isgur and M.B. Wise, Phys. Lett.  {\bf 232B} (1989) 113;
            Phys. Lett.  {\bf 237B} (1990) 527;

\bibitem{2}  H. Georgi, Phys. Lett.  {\bf 264B} (1991) 447.

\bibitem{3}  H. Georgi, B. Grinstein, and M.B. Wise,
            Phys. Lett.  {\bf 252B} (1990) 456.

\bibitem{4}  M.E. Luke, Phys. Lett.  {\bf 252B} (1990) 447.

\bibitem{5}  A.F. Falk, B. Grinstein, and M.E. Luke,
             Harvard preprint HUTP-90/A044  (1990).

\bibitem{6} E. Eichten and B. Hill,  Phys. Lett.  {\bf 234B} (1990) 511.


\bibitem{7}  C.-H. Chang and Y.-Q. Chen, Phys. Rev. D {\bf 46 } (1992) 3845;
             Phys. Lett. {\bf 284 } 127.

\bibitem{8}  C.-H. Chang and Y.-Q. Chen, Phys. Rev. D {\bf 48 } (1993)  .

\bibitem{9}  C.-H. Chang and Y.-Q. Chen, preprint AS-ITP-92-75, CCAST-92-41
           {\it The Decays of $B_c$ Meson} to be published in Phys. Rev. D.

\bibitem{10} C.-R. Ji and F. Amiri, Phys. Rev.{\bf D35} (1987) 3318;
             X. Artru and G. Mennessier, Nucl. Phys. {\bf
             70} (1974) 93.

\bibitem{11} B. Andersson, G.Gustafson, G. Ingelman
             and T.Sj\"{o}strand, Phys. Rep. {\bf 97} (1983) 31;
             S.S. Gershtein, A.K. Likhoded and S.R. Slabospitsky, Intern'l.
             J. Mod. Phys. {\bf 6} (1991) 2309; V.V. Kiselev et al., Sov. J.
             Nucl. Phys. {\bf 49} (1989) 682.

\bibitem{12} CDF Collaboration, F. Abe {\it et al.} Phys. Rev. Lett. {\bf68}
             (1992) 448.

\bibitem{13} S. Mandelstam, Proc. Roy. Soc. {\bf233} (1955) 248.

\bibitem{14} E. E. Salpeter, Phys. Rev. {\bf 87} (1952) 328.

\bibitem{15} C.-H. Chang and T.-H. Ho, High Energy Physics and Nuclear
             Physics {\bf 2} (1978) 119 ({\it IN CHINESE}).

\bibitem{16} M. Wirbel, B. Stech, and M. Bauer, Z. Phys. {\bf C29} (1985) 637;
             M. Wirbel, B. Stech, and M. Bauer, Z. Phys. {\bf C34} (1987) 103.

\bibitem{17}  B. Grinstein, M.B. Wise, and N. Isgur,
           Phys.  Rev.  Lett.  {\bf 56} (1986) 298;
           N. Isgur, D. Scora, B. Grinstein, M.B. Wise,
           Phys.  Rev. D  {\bf 39} (1989) 799.

\bibitem{18} B. Durand and L. Durand Phys. Rev. D {\bf 25} (1982) 2312.

\bibitem{19} R.Kleiss and W.J.Stirling, Nucl. Phys.{\bf B262} (1985) 235.

\bibitem{20} T.Altomari and L. Wolfenstein, Phys. Rev. Lett. {\bf 58}
             (1987) 1563;
          T. Altomari and L. Wolfenstein, Phys. Rev. D {\bf 37} (1988) 681.

\bibitem{21} W.  Buchm\"{u}ller, G. Grunberg and
             S. -H. H. Tye, Phys.  Rev.  Lett.  {\bf 45} (1980) 103;
             W.  Buchm\"{u}ller and S. -H. H. Tye Phys.  Rev.  D
             {\bf 24} (1981) 132.

\bibitem{22}  M. Lusignoli and M. Masetti, Z. Phys. C {\bf 51} (1991) 459.

\bibitem{23} Y.-Q. Chen and Y.-P. Kuang, Phys. Rev. D {\bf 46} (1992) 1165.

\bibitem{24} J.D. Bjorken, invited talk given at Les Recontre de la
             Valle d'Aoste La Thuile, Aosta Valley, Italy, March 1990,
             SLAC report SLAC-PUB-5278 (1990); J.D. Bjorken, I. Dunietz,
             and J.Taron, Nucl. Phys. {\bf B371} (1992) 111.

\bibitem{25} N. Cabibbo and L. Radicati, Phys. Lett.  {\bf 19} (1966) 697.

\bibitem{26}  C.-H. Chang and Y.-Q. Chen, in preparation.

\end{thebibliography}
\end{document}